\documentclass[12pt]{article}
\usepackage{epsfig,rotating,setspace,latexsym,amsmath,epsf,amssymb,bm}
\usepackage{cite}

\newtheorem{theorem}{Theorem}
\newtheorem{corollary}{Corollary}
\newtheorem{lemma}{Lemma}

\begin{document}

\title{Capacity Regions and Bounds for a Class of Z-interference Channels
\thanks{This work was
supported by the DARPA ITMANET program under grant 1105741-1-TFIND. 
}
}

\author{Nan Liu \qquad Andrea Goldsmith \\
\normalsize Department of Electrical Engineering \\
\normalsize Stanford University, Stanford CA 94305 \\
\normalsize {\it nanliu@stanford.edu} \qquad {\it andrea@wsl.stanford.edu} }
\date{}
\maketitle

\begin{abstract}
We define a class of Z-interference channels for which we obtain a new upper bound on the capacity region. The bound exploits a technique first introduced by Korner and Marton. A channel in this class has the property that, for the transmitter-receiver pair that suffers from interference, the conditional output entropy at the receiver is invariant with respect to the transmitted codewords. We compare the new capacity region upper bound with the Han/Kobayashi achievable rate region for interference channels. This comparison shows that our bound is tight in some cases, thereby yielding specific points on the capacity region as well as  sum capacity for certain Z-interference channels. In particular, this result can be used as an alternate method to obtain sum capacity of Gaussian Z-interference channels. We then apply an additional restriction on our channel class: the transmitter-receiver pair that suffers from interference achieves its maximum output entropy with a single input distribution irrespective of the interference distribution. For these channels we show that our new capacity region upper bound coincides with the Han/Kobayashi achievable rate region, which is therefore capacity-achieving. In particular, for these channels superposition encoding with partial decoding is shown to be optimal and a single-letter characterization for the capacity region is obtained.
\end{abstract}

\vspace*{0.4cm}

\newpage

\section{Introduction}
The interference channel,
introduced in \cite{Shannon:1961}, is a simple network
consisting of two pairs of transmitters and receivers. Each pair
wishes to communicate at a certain rate with negligible
probability of error. However, the two communications interfere
with each other. The problem of finding the capacity region of the interference channel is difficult and therefore remains
essentially open except in some special cases, e.g., interference channels with statistically equivalent
outputs \cite{Ahlswede:1971,Sato:1977,Carleial:1978}, discrete additive degraded
interference channels \cite{Benzel:1979}, a class of
deterministic interference channels \cite{ElGamal:1982}, strong interference channels
\cite{Carleial:1975, Sato: 1978MayIC, Han: 1981, Sato:1981,Costa:1987}, and a class of degraded interference channels \cite{Liu_Ulukus:2006Allerton}.

While the capacity region of the interference channel remains largely unknown, achievable rate regions and capacity region upper bounds have been found. To date, techniques for obtaining achievable rate regions for the interference channel include treating interference as noise \cite{Ahlswede: 1971}, fully decoding the unwanted interference \cite{Sato:1977}, superposition encoding and partially decoding the unwanted interference \cite{Carleial:1978, Han:1981}, and time division multiplexing/frequency division multiplexing \cite{Sato:1977,Carleial:1978}. Techniques for upper bounding the capacity region of the interference channel include allowing transmitters and receivers to cooperate without requiring additional bandwidth or power \cite{Sato:1977}, introducing an imaginary but more capable receiver into the system \cite{Carleial: 1983}, and genie-aided receivers \cite{ElGamal:1982, Kramer:2004, Etkin: 2007, Telatar:2007}.

The Z-interference channel is an interference channel where one transmitter-receiver pair is interference-free. Though finding the capacity region of the Z-interference channel is a simpler problem than that of the interference channel, capacity results are still limited, with the following exceptions: The Z-interference channel capacity region is known when the interference is deterministic \cite[Section IV]{ElGamal:1982}. The Z-interference channel sum capacity is known when the Z-interference channel is Gaussian \cite{Sason:2004}, or when the interference-free link is also noise-free \cite{Ahlswede:2006}, although in both cases, the full capacity region has not been characterized.

In this paper, we first provide a new capacity region upper bound for a class of Z-interference channels that satisfy a given condition: namely that the performance of the transmitter-receiver pair that suffers from interference depends only on the distance of the transmitter's codewords and not their exact locations. The technique \cite[page 314]{Csiszar:book} that we use in obtaining the converse  associated with this upper bound was introduced by Korner and Marton in \cite{Korner:1977Image}. This technique has been useful in the solution of several problems in multi-user information theory, in particular for broadcast channels with degraded message sets \cite{Korner:1977}, communication where the transmitter has non-causal perfect side information, i.e., the Gelfand-Pinsker problem \cite{Gelfand:1980}, and semicodes for the multiple access channel \cite{Ahlswede:2006}. We then compare the new capacity region upper bound with the Han/Kobayashi achievable rate region \cite{Han:1981}. This comparison yields certain points on the capacity region, including the sum-rate point, for channels in our defined class.

We next add an  additional condition to our channel class, namely that for the transmitter-receiver pair that suffers from interference, the maximum output entropy can always be achieved regardless of the choice of the interferer's codebook. With this additional condition we show that our capacity region upper bound coincides with the Han/Kobayashi achievable rate region, and hence equals the capacity region for this subclass of Z-interference channels. The Han/Kobayashi achievable rate region is obtained using the idea of superposition encoding and partial decoding.  Specifically, the transmitters are required to encode their messages via superposition encoding, and each receiver is required to decode not only its own message, but also part of the interference. One of the main difficulties in finding the capacity region of the interference channel is to justify the need for partial decoding in the converse proofs, as the partially decoded information is not required at the receiver. Using our new capacity region upper bound, we are able justify the optimality of superposition encoding and partial decoding for a subclass of Z-interference channels. So far, the only result that proves superposition encoding and partial decoding is optimal is \cite{ElGamal:1982} where, due to the deterministic nature of the channel, the capacity region upper bound obtained by using the technique of genie-aided receivers is sufficient to meet the achievable rate region of \cite{Han:1981}.
Using a different upper bounding technique, we show that superposition encoding and partial decoding is optimal for certain Z-interference channels, which are not necessarily deterministic and, therefore, provide a single-letter characterization for the capacity
region of these channels, which was previously unknown.

The remainder of this paper is organized as follows. In Section \ref{system_model}, we define the class of Z-interference channels investigated in this paper by providing two conditions on the channels.  In Section \ref{upper_bound}, we provide a new capacity region upper bound for Z-interference channels satisfying the first conditon. We restate the  Han/Kobayashi achievable rate region for Z-interference channels in Section \ref{lower_bound}. In Section \ref{compcomp}, points on the capacity region as well as sum capacity for certain Z-interference channels are found by comparing our capacity region upper bound with the Han/Kobayashi achievable rate region. In Section \ref{capacity_region}, we provide the single-letter characterization of the capacity region for Z-interference channels that satisfy both conditions. This is followed by our conclusions in Section \ref{conclusions}. Details of certain proofs are given in the Appendix in Section \ref{appendix}.

\section{System Model} \label{system_model}
Consider a Z-interference channel with two transition probabilities $p(y_1|x_1)$ and
$p(y_2|x_1,x_2)$. The input and output alphabets are
$\mathcal{X}_1$, $\mathcal{X}_2$, $\mathcal{Y}_1$ and
$\mathcal{Y}_2$. Set
\begin{align}
V_1(a|b)&=\text{Pr}[Y_1=a|X_1=b],\\
V_2(c|b,d)&=\text{Pr}[Y_2=c|X_1=b,X_2=d].
\end{align}
Let $W_1$ and $W_2$ be two independent messages uniformly distributed on $\{1,2,\cdots,M_1\}$ and $\{1,2,\cdots,M_2\}$, respectively.
Transmitter $i$ wishes to send message $W_i$ to Receiver $i$, $i=1,2$. An $(M_1,M_2,n,\epsilon_n)$ code for this channel consists of a sequence of two encoding functions
\begin{align}
f_i^n: \{1,2,\cdots,2^{n R_i}\} \rightarrow \mathcal{X}_i^n, \qquad i=1,2
\end{align}
and two decoding functions
\begin{align}
g_i^n: \mathcal{Y}_i^n \rightarrow \{1,2,\cdots,2^{n R_i}\}, \qquad i=1,2
\end{align}
with probability of error 
\begin{align}
\epsilon_n=\max_{i=1,2} \quad \frac{1}{M_1 M_2} \sum_{w_1,w_2} \text{Pr} \left[g_i^n(Y_i^n) \neq w_i|W_1=w_1,W_2=w_2 \right].
\end{align}
A rate pair $(R_1,R_2)$ is said to be achievable if there exists a sequence of $\left(2^{nR_1}, 2^{nR_2}, n, \epsilon_n \right)$ codes such that $\epsilon_n \rightarrow 0$ as $n \rightarrow \infty$. The capacity region of the Z-interference channel is the closure of the set of all achievable rate pairs. 

An example of the Z-interference channel is the Gaussian Z-interference channel, where $\mathcal{X}_1=\mathcal{X}_2=\mathcal{Y}_1=\mathcal{Y}_2=\mathbb{R}$, and $p(y_1|x_1)$ and $p(y_2|x_1,x_2)$ are given as
\begin{align}
Y_1&=X_1+Z_1, \label{Gauss1}\\
Y_2&=a X_1+X_2+Z_2, \label{Gauss2}
\end{align}
where $Z_1$ and $Z_2$ are independent Gaussian random variables with zero mean and unit variance, $a \in \mathbb{R}$, and the channel inputs have to satisfy the average power constraints of $P_1$ and $P_2$.

The class we investigate in this paper consists of Z-interference channels that satisfy :

\vspace{0.2in}

\noindent
{\bf{Condition 1}}:
For any $n=1,2,\cdots$, $H(Y_2^n|X_2^n=x_2^n)$, when evaluated with the distribution $\sum_{x_1^n}p(x_1^n)p(y_2^n|x_1^n,x_2^n)$,  is independent of $x_2^n$ for any $p(x_1^n)$.

\vspace{0.2in}

Condition 1 specifies that the channel $p(y_2|x_1,x_2)$ is invariant, in terms of conditional output entropy, with respect to the input sequence of Transmitter 2, i.e., $x_2^n$. This means that when designing the codebook of Transmitter 2, the exact locations of the codewords do not affect the performance, rather, it is the relative locations, or ``distances'', between codewords that matter. For example, the Gaussian Z-interference channel, defined in (\ref{Gauss1}) and (\ref{Gauss2}),  satisfies this condition. 

Define $\tau$ as
\begin{align}
\tau=\max_{p(x_1)p(x_2)} H(Y_2). \label{definetau}
\end{align}
The class of Z-interference channels for which we are able to obtain the capacity region satisfy Condition 1 as well as the following condition. 

\vspace{0.2in}

\noindent
{\bf{Condition 2}}: There exists a $p^*(x_2)$ such that $H(Y_2)$, when evaluated with the distribution $\sum_{x_1,x_2} p(x_1)p^*(x_2)p(y_2|x_1,x_2)$, is equal to $\tau$ for any $p(x_1)$.

\vspace{0.2in}

Intuitively, Condition 2 specifies that no matter how tightly packed the codewords in codebook 1 are, by spacing out the codewords in codebook 2, we can always fill up the entire, or maximum, output space at Receiver 2. This means that using an i.i.d. generated codebook with $p^*(x_2)$ at Transmitter 2 is to our advantage, as the larger the output space, the more codewords of Transmitter 2 we can pack in the space. Note that the Gaussian Z-interference channel does not satisfy this condition, since the largest output space is only achieved when both $p(x_1)$ and $p(x_2)$ are Gaussian with variances $P_1$ and $P_2$, respectively. The largest output space cannot be achieved with a $p^*(x_2)$ that is irrespective of $p(x_1)$, as specified in Condition 2.

Now, we give an example of a channel, shown in Figure \ref{eg}, where both conditions are satisfied. Let $\mathcal{X}_2=\mathcal{Y}_2=\mathcal{S}=\{0, 1, 2, \cdots, q-1\}$, where $q$ is an arbitrary interger. Let sets $\mathcal{X}_1$, $\mathcal{Y}_1$, and probability distributions $\bar{p}(y_1|x_1)$ and $\bar{p}(s|x_1)$ be arbitrary. The channel is defined as $V_1(y_1|x_1)=\bar{p}(y_1|x_1)$, and $V_2(y_2|x_1,x_2)$ is defined as $\sum_{s \in \mathcal{S}} \bar{p}(y_2|s, x_2)\bar{p}(s|x_1)$, where $\bar{p}(y_2|s, x_2)$ is given by
\begin{align}
Y_2=S \oplus X_2. \label{andrea}
\end{align}
In (\ref{andrea}), $\oplus$ is the mod-$q$ sum. It is easy to see that the channel thus defined satisfies Condition 1. By letting $p^*(x_2)$ be the uniform distribution on $\{0, 1, 2, \cdots, q-1\}$, we find that the channel also satisfies Condition 2, where $\tau=\log |\mathcal{Y}_2|=\log q$.

The above example is related to several Z-interference channels that have been studied in \cite{Benzel:1979, ElGamal:1982, Telatar:2007}, as we now describe in more detail.
\begin{enumerate} 
\item The discrete additive degraded
interference channels studied in \cite{Benzel:1979}, using similar techniques as \cite[Fig. 6]{Costa:1985}, can be shown to be equivalent to, or in other words, have the same capacity region as, the following Z-inteference channel:
\begin{align}
Y_1&=X_1 \oplus Z_1 \label{benzel1}\\
Y_2&=X_1 \oplus X_2 \oplus Z_1 \oplus Z_2. \label{benzel2}
\end{align}
The Z-interference channel characterized by (\ref{benzel1}) and (\ref{benzel2}) is a special case of our example.
The derivation of the capacity region in \cite{Benzel:1979} relies on the degradedness of output $Y_2$ with respect to $Y_1$, which makes treating interference as noise optimal. In our example, we do not make  an assumption on degradedness, and show that superposition encoding and partial decoding is optimal.
\begin{figure}
\centering
\includegraphics[width=3.5in]{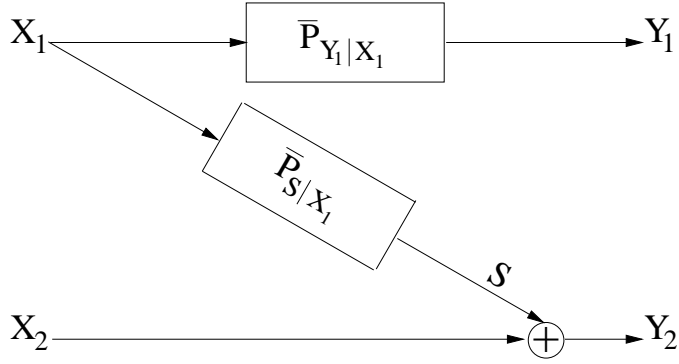}
\caption{Example of a Z-interference channel satisfying Conditions 1 and 2}\label{eg}
\end{figure}

\item The example shown in \cite[Fig. 3]{ElGamal:1982} is a special case of our example, with $\mathcal{X}_2=\mathcal{Y}_2=\{0,1\}=\mathcal{S}=\mathcal{Y}_1$ and $\mathcal{X}_1=\{0,1,2\}$, $\bar{p}(y_1|x_1)$ is given as
\begin{align}
\bar{p}(y_1|x_1)=\left\{ 
\begin{array}{ll}
1& \text{ if } (x_1,y_1)=(0,0) \text{ or } (x_1,y_1)=(1,1)\\
\epsilon & \text{ if } (x_1,y_1)=(2,0)\\
1-\epsilon & \text{ if } (x_1,y_1)=(2,1)\\
0 & \text{ otherwise }
\end{array} \right.
\end{align}
and $\bar{p}(s|x_1)$ is given as
\begin{align}
\bar{p}(s|x_1)=\left\{ 
\begin{array}{ll}
1& \text{ if } (x_1,s)=(0,0) \text{ or } (x_1,s)=(1,1) \text{ or } (x_1,s)=(2,0)\\
0 & \text{ otherwise }
\end{array} \right.
\end{align}
which is deterministic. As mentioned before, the capacity region of this channel is found in \cite{ElGamal:1982} by matching the achievable rate region of superposition encoding and partial decoding and the capacity region upper bound of genie-aided receivers. The deterministic nature of $p(y_2|x_1,x_2)$ plays an important role in obtaining the results. In our example, we make no assumption on the channel being deterministic, and use a different upper bounding technique to match the achievable rate region of superposition encoding and partial decoding.

\item For a class of interference channels, \cite{Telatar:2007} quantifies the gap between the achievable rate region of superposition encoding and partial decoding and the capacity region upper bound obtained using the technique of genie-aided receivers. Consider a special case of the class of interference channels considered in \cite{Telatar:2007}, where $p(s_2|x_2)$ in \cite[Fig. 1]{Telatar:2007} is independent of $x_2$ and the deterministic function $f_2$ in \cite[Fig. 1]{Telatar:2007} is the modulo sum operation. This special case is contained in our example. Hence, for this special case, the results in this paper provide the exact capacity region, and we may conclude that the achievable rate region of Han and Kobayashi used in \cite{Telatar:2007} is in fact optimal, while the capacity region upper bound in \cite{Telatar:2007} is not tight.
\end{enumerate}

As a final remark in this section, note that in general, Condition 1 can not be verified in a ``single-letter'' way, i.e., by checking for only the case of $n=1$. However, for many channels, such as the Gaussian Z-interference channel in (\ref{Gauss1}) and (\ref{Gauss2}) and the example given in Figure \ref{eg}, Condition 1 is straightforward to verify.

\section{New Capacity Region Upper Bounds} \label{upper_bound}
In this section, we obtain a new capacity region upper bound for the class of Z-interference channels that satisfy Condition 1.

Before we proceed, we will first restate the converse technique introduced in Korner and Marton \cite{Korner:1977Image}. The technique is a method of writing the difference between two $n$-letter entropies into a sum of differences between conditional entropies, which may then be written as the difference between two single letter conditional entropies by defining the appropriate auxiliary random variables.

\begin{lemma} \cite[page 314, eqn (3.34)]{Csiszar:book} \label{nletterdifference}

For any $n$, and any random variables $Y^n$ and $Z^n$, we have 
\begin{align}
H(Z^n)-H(Y^n)=\sum_{i=1}^n \left(H(Z_i|Y^{i-1}, Z_{i+1}, Z_{i+2}, \cdots, Z_n)-H(Y_i|Y^{i-1}, Z_{i+1}, Z_{i+2}, \cdots, Z_n) \right).
\end{align}
\end{lemma}

Now, we state our upper bound for the class of Z-interference channels that satisfy Condition 1.
\begin{theorem} \label{converse}
For a Z-interference channel, characterized by transition probabilities $V_1$ and $V_2$, that satisfies Condition 1, if rate pair $(R_1,R_2)$ is achievable, then it must satisfy
\begin{align}
R_1 & \leq H(Y_1|U, Q)+\gamma-H(Y_1|X_1) \label{upper1}\\
R_2 &\leq H(Y_2|Q)-H(T|U, Q)-\gamma \label{upper2}\\
0 \leq \gamma &\leq \min(I(Y_1;U|Q), I(T;U|Q)) \label{upper3}
\end{align}
for some distribution $p(q)p(x_1,u|q)p(x_2|q)$ and number $\gamma$, where the mutual informations and entropies are evaluated using
$p(q, u,x_1, x_2, y_1,t, y_2)=p(q)p(x_1,u|q)p(x_2|q) V_1(y_1|x_1)V_2(t|x_1,\bar{x}_2) \break V_2(y_2|x_1,x_2)$, and $\bar{x}_2$ is an arbitrary element in $\mathcal{X}_2$.
\end{theorem}
Theorem \ref{converse} is the major technical contribution of this work. However, to improve the readability of the paper, we have moved the proof of Theorem \ref{converse} to the Appendix in Section \ref{converseproof}. The key point of the proof is to define an imaginary memoryless channel $\hat{V}_2: \mathcal{X}_1 \rightarrow \mathcal{Y}_2$, with input $X_1$ and output $T$, as
\begin{align}
\hat{V}_2(t|x_1)=V_2(t|x_1, \bar{x}_2),
\end{align} 
where $\bar{x}_2$ is an arbitrary element in $\mathcal{X}_2$. Theorem \ref{converse} is proved by using the definition of this imaginary channel, Lemma \ref{nletterdifference} and Condition 1. Notice that $T$ is a random variable that we created in the proof of Theorem \ref{converse} which does not exist in the actual communication system.

\section{Achievable rates for Z-interference channels} \label{lower_bound}
In this section, we specialize the achievable rate region in \cite{Han:1981}, which is developed for the interference channel, to the Z-interference channel. More specifically, we use, as in \cite{Chong:2006}, Fourier-Motzkin elimination and the fact that $p(u_i|q)p(w_i|q)f(x_i|u_i,w_i,q)$ is sufficient to achieve all possible marginals $p(w_i|q)p(x_i|w_i,q)$ for $i=1,2$. By setting the auxiliary random variable associated with Receiver 2 to be constant, we obtain the following achievable rate region.

Define $\mathcal{P}$ as the set of distributions $p(q,u,x_1,x_2)$ that satisfies
\begin{align}
p(q,u,x_1,x_2)=p(q)p(x_1,u|q)p(x_2|q).
\end{align}
For each $p \in \mathcal{P}$, further define the region $\mathcal{G}_I(p)$ as
\begin{align}
\mathcal{G}_I(p) \overset{\triangle}{=} \{(R_1,R_2)|& R_1  \leq I(X_1;Y_1|U,Q)+\min(I(Y_1;U|Q),
 I(Y_2;U|X_2, Q)), \nonumber\\
& R_2 \leq I(X_2;Y_2|U, Q), \nonumber\\
& R_1+R_2  \leq I(X_1;Y_1|U,Q)+ I(U, X_2;Y_2|Q) \}.
\end{align}
where the mutual informations are evaluated with $p$, and the given channel transition probabilities $p(y_1|x_1)$ and $p(y_2|x_1,x_2)$.

\begin{theorem}\cite{Han: 1981, Chong:2006} \label{achach}
For the Z-interference channel described by $V_1$ and $V_2$, an inner bound on the capacity region is
\begin{align}
\mathcal{G}_I=\bigcup_{p \in \mathcal{P}} \mathcal{G}_I(p).
\end{align}
Furthermore, the inner bound remains invariant if we impose the following constraint on the cardinality of the auxiliary random variables:
\begin{align}
|\mathcal{U}| \leq |\mathcal{X}_1|+2, \quad |\mathcal{Q}| \leq 4.
\end{align}
\end{theorem}

\section{Comparison of achievable rates and rate upper bounds} \label{compcomp}
Since $T$ is a random variable that we created in the proof of Theorem \ref{converse} which does not exist in Theorem \ref{achach}, in order to compare the new capacity region upper bound with the known achievable rate region, we rewrite the upper bound in a different way by replacing $T$ with $Y_2$ conditioned on $X_2$, using Condition 1. In other words, the capacity region upper bound given in Theorem \ref{converse} is equivalent to the one given in the next theorem.
\begin{theorem} \label{converse2}
For a Z-interference channel, characterized by transition probabilities $V_1$ and $V_2$, that satisfies Condition 1, if the rate pair $(R_1,R_2)$ is achievable, then it must satisfy
\begin{align}
R_1 & \leq I(X_1;Y_1|U,Q)+\gamma \label{upper11}\\
R_2 &\leq I(U, X_2;Y_2|Q)-\gamma \label{upper22}\\
0 \leq \gamma &\leq \min(I(Y_1;U|Q), I(Y_2;U|X_2, Q)) \label{upper33}
\end{align}
for some distribution $p(q)p(x_1,u|q)p(x_2|q)$ and number $\gamma$, where the mutual informations are evaluated using
$p(q, u,x_1, x_2, y_1, y_2)=p(q)p(x_1,u|q)p(x_2|q) V_1(y_1|x_1) V_2(y_2|x_1,x_2)$.
\end{theorem}
A proof of Theorem \ref{converse2} is provided in the Appendix in Section \ref{conversedifferent}. Based on Theorem \ref{converse2}, we now restate the capacity region upper bound as follows, by eliminating variable $\gamma$. The proof is straightforward, and thus omitted. For each $p \in \mathcal{P}$, define the region $\mathcal{G}_o(p)$ as
\begin{align}
\mathcal{G}_o (p) \overset{\triangle}{=} \{(R_1,R_2)|& R_1  \leq I(X_1;Y_1|U,Q)+\min(I(Y_1;U|Q),
 I(Y_2;U|X_2, Q)), \nonumber\\
& R_2 \leq I(U, X_2;Y_2|Q), \nonumber\\
& R_1+R_2  \leq I(X_1;Y_1|U,Q)+ I(U, X_2;Y_2|Q) \}.
\end{align}
where the mutual informations are evaluated with $p$, and the given channel transition probabilities $p(y_1|x_1)$ and $p(y_2|x_1,x_2)$.

\begin{theorem} \label{realconverse}
For a Z-interference channel, characterized by transition probabilities $V_1$ and $V_2$, that satisfies Condition 1, an outer bound on the capacity region is
\begin{align}
\mathcal{G}_o=\bigcup_{p \in \mathcal{P}} \mathcal{G}_o(P).
\end{align}
Furthermore, the outer bound remains invariant if we impose the following constraint on the cardinality of auxiliary random random variables:
\begin{align}
|\mathcal{U}| \leq |\mathcal{X}_1|+1, \quad |\mathcal{Q}| \leq 4.
\end{align}
\end{theorem}

\begin{figure}
\centering
\includegraphics[width=3.5in]{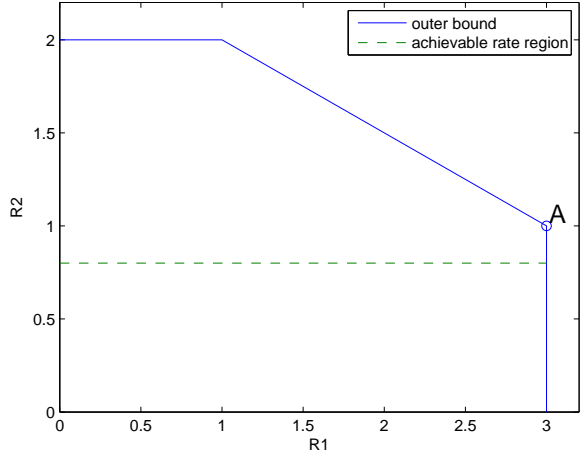}
\caption{Comparison of $\mathcal{G}_I(p)$ and $\mathcal{G}_o(p)$ when $I(U;Y_1|Q) < I(U;Y_2|Q)$}\label{case1}
\end{figure}

\begin{figure}
\centering
\includegraphics[width=3.5in]{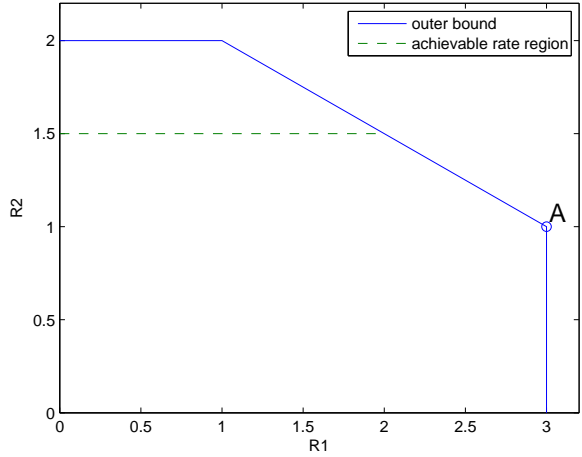}
\caption{Comparison of $\mathcal{G}_I(p)$ and $\mathcal{G}_o(p)$ when $I(U;Y_1|Q) \geq I(U;Y_2|Q)$}\label{case2}
\end{figure}

Comparing the capacity region upper bound in Theorem \ref{realconverse} and the achievable rate region in Theorem \ref{achach}, we see that the set of allowable distributions, $\mathcal{P}$, are the same, while the regions, $\mathcal{G}_I(p)$ and $\mathcal{G}_o(p)$, look slightly different for the same $p \in \mathcal{P}$. More specifically, among the three equations that characterize $\mathcal{G}_I(p)$ and $\mathcal{G}_o(p)$, two of them are exactly the same. Notice also the fact that, the region $\mathcal{G}_o(p)$ is always a pentagon, while the region $\mathcal{G}_I(p)$ can be a pentagon if $p$ satisfies $I(U;Y_1|Q) \geq I(U;Y_2|Q)$, and a rectangle otherwise. We illustrate the gap between $\mathcal{G}_I(p)$ and $\mathcal{G}_o(p)$ in the following two figures. If the distribution $p$ is such that $I(U;Y_1|Q) < I(U;Y_2|Q)$, the comparison between $\mathcal{G}_I(p)$ and $\mathcal{G}_o(p)$ looks like Figure \ref{case1}. On the other hand, if the distribution $p$ is such that $I(U;Y_1|Q) \geq I(U;Y_2|Q)$, the comparison would look like Figure \ref{case2}. In this case, corner point $A$ is achievable.

The outer bound, $\mathcal{G}_o$, is a union of pentagons, and the boundary of $\mathcal{G}_o$ consists of the two corner points of certain pentagons. It is likely that point $A$ of $\mathcal{G}_o(p_0)$ for some $p_0 \in \mathcal{P}$, will appear on the boundary of $\mathcal{G}_o$. If $p_0$ also satisfies $I(U;Y_1|Q) \geq I(U;Y_2|Q)$, then point $A$ is a \emph{capacity} point, i.e., a point on the boundary of the capacity region. We formalize this fact in Corollary \ref{capacitypoint}.
\begin{corollary} \label{capacitypoint}
For Z-interference channels that satisfy Condition 1,
if point $A$ given by: 
\begin{align}
\big(I(X_1;Y_1|U,Q)&+\min(I(Y_1;U|Q),
 I(Y_2;U|X_2, Q)), \nonumber\\
 &I(U, X_2;Y_2|Q) -\min(I(Y_1;U|Q),
 I(Y_2;U|X_2, Q)) \big)
\end{align}
for some distribution $p \in \mathcal{P}$ that satisfies $I(U;Y_1|Q) \geq I(U;Y_2|Q)$ happens to be on the boundary of $\mathcal{G}_o$, then point $A$ is a point on the boundary of the capacity region.
\end{corollary}
Further, by comparing the achievable rate region and capacity region upper bound, we may make the following statement about the sum capacity of Z-interference channels that satisfy Condition 1.
\begin{corollary} \label{corosum}
For Z-interference channels that satisfy Condition 1, if
\begin{align}
\max_{p(q)p(x_1,u|q)p(x_2|q)} I(X_1;Y_1|U,Q)+ I(U, X_2;Y_2|Q) \label{sumcapacity}
\end{align}
is achieved by a distribution that satisfies $I(U;Y_1|Q) \geq I(U;Y_2|Q)$, then the sum capacity is (\ref{sumcapacity}).
\end{corollary}
To demonstrate how Corollary \ref{corosum} might be used, consider the Gaussian Z-interference channel, defined in (\ref{Gauss1}) and (\ref{Gauss2}), where $a \leq 1$. For any $p(q)p(x_1,u|q)p(x_2|q)$, $I(U;Y_1|Q) \geq I(U;Y_2|Q)$ is satisfied. Hence, (\ref{sumcapacity}) is the sum capacity. The sum capacity of the Gaussian Z-interference channel where $a \leq 1$ has been found in \cite{Sason:2004} using a different technique.

\section{Capacity region of a subclass of Z-interference channels} \label{capacity_region}
As illustrated in Section \ref{compcomp},
in general, for Z-interference channels that satisfy Condition 1, the capacity region upper bound, described in Theorem \ref{realconverse}, and the achievable rate region, described in Theorem \ref{achach}, are not the same. However, in this section we show that when the Z-interference channel satisfies the additional Condition 2, our capacity region upper bound and the Han/Kobayashi achievable rate region coincide, yielding the capacity region.

In the achievable rate region derived in Theorem \ref{achach}, if we specify $Q$ to be a constant, i.e., time sharing is not employed, and if we further specify $p(x_2)$ to be $p^*(x_2)$, defined in Condition 2, we obtain another achievable rate region:
\begin{align}
R_1 &\leq I(X_1;Y_1|U)+\min (I(U;Y_1), I(U;Y_2|X_2)) \label{achcon1} \\
R_2 &\leq \tau-H(Y_2|X_2,U) \label{achcon2}\\
R_1 + R_2 & \leq I(X_1;Y_1|U)+\tau-H(Y_2|X_2,U) \label{achcon3}
\end{align}
for some distribution $p(u)p(x_1|u)$,
where the mutual informations and entropies are evaluated with 
\begin{align}
p(u,x_1,x_2,y_1,y_2)=p(u)p(x_1|u)p^*(x_2)V_1(y_1|x_1)V_2(y_2|x_1,x_2). \label{achdis}
\end{align}
In obtaining this region, we have used the fact that the Z-interference channel satisfies Condition 2, and therefore, we have
\begin{align}
H(Y_2|U)=\sum_{u} p(u) H(Y_2|U=u) = \tau=H(Y_2).
\end{align}
This achievable rate region is potentially smaller than that described in Theorem \ref{achach}, due to the fact that we have eliminated time sharing and fixed $p(x_2)$ to be a specific distribution.

As for the upper bound, rather than starting from the capacity region upper bound described in Theorem \ref{realconverse}, we study the equivalent region described in Theorem \ref{converse}. By the definition of $\tau$ in (\ref{definetau}), another upper bound on the Z-interference channel that satisfies Condition 1, which is looser than that of Theorem \ref{converse}, is
\begin{align}
R_1 & \leq H(Y_1|U, Q)+\gamma-H(Y_1|X_1) \label{looserouter1} \\
R_2 &\leq \tau-H(T|U, Q)-\gamma \label{looserouter2}\\
0 \leq \gamma &\leq \min(I(Y_1;U|Q), I(T;U|Q))  \label{looserouter3}
\end{align}
for some distribution $p(q,u)p(x_1|u,q)$ and number $\gamma$, where the mutual informations and entropies are evaluated using
$p(q, u,x_1, y_1,t)=p(q, u)p(x_1|u,q) V_1(y_1|x_1)V_2(t|x_1,\bar{x}_2)$. Note that by replacing $H(Y_2|Q)$, or more specifically $H(Y_2|Q=q)$, with a possibly larger term $\tau$, we have removed the dependence of the region on $p(x_2|q)$.

By defining a new auxiliary random variable $\bar{U}$ to be $(U,Q)$, and noting the fact that 
\begin{align}
H(Y_1|Q) &\leq H(Y_1)\\
H(T|Q) & \leq H(T),
\end{align}
we obtain yet another capacity region upper bound, which is possibly looser than the previous one described in (\ref{looserouter1})-(\ref{looserouter3}):
\begin{align}
R_1 & \leq H(Y_1|\bar{U})+\gamma-H(Y_1|X_1) \label{what1}\\
R_2 &\leq \tau-H(T|\bar{U})-\gamma \\
0 \leq \gamma &\leq \min(I(Y_1;\bar{U}), I(T;\bar{U})) \label{what3}
\end{align}
for some distribution $p(\bar{u})p(x_1|\bar{u})$ and number $\gamma$, where the mutual informations and entropies are evaluated using
$p(\bar{u},x_1, y_1,t)=p(\bar{u})p(x_1|\bar{u}) V_1(y_1|x_1)V_2(t|x_1,\bar{x}_2)$. By defining $\bar{U}$ to be $(U,Q)$, we have gotten rid of the time sharing auxiliary random variable $Q$.

For notational convenience, we replace $\bar{U}$ with $U$, and eliminate variable $\gamma$, to get the following capacity region upper bound for Z-interference channels satisfying Condition 1, which is equivalent to the capacity region upper bound described in (\ref{what1})-(\ref{what3}):
\begin{align}
R_1 & \leq I(Y_1;X_1|U)+\min(I(Y_1;U), I(T;U)) \label{specialupper1}\\
R_2 &\leq \tau-H(T|U) \label{specialupper2}\\
R_1+R_2 & \leq I(Y_1;X_1|U)+\tau-H(T|U) \label{specialupper3}
\end{align}
for some distribution $p(u)p(x_1|u)$, where the mutual informations and entropies are evaluated using
\begin{align}
p(u,x_1, y_1,t)=p(u)p(x_1|u) V_1(y_1|x_1)V_2(t|x_1,\bar{x}_2). \label{convdis}
\end{align}

Next, we will show that when the Z-interference channel satisfies Conditions 1 and 2, the capacity region upper bound described by (\ref{specialupper1})-(\ref{specialupper3}) and the achievable rate region described by (\ref{achcon1})-(\ref{achcon3}) are the same when evaluated with the same $p(u)p(x_1|u)$.

When evaluated with the same $p(u)p(x_1|u)$, it is obvious that the terms $H(Y_1|U)$, $H(Y_1|X_1)$ and $I(U;Y_1)$ are equal in both (\ref{specialupper1})-(\ref{specialupper3}) and (\ref{achcon1})-(\ref{achcon3}). $H(T|U)$, respectively $H(T)$, evaluated with the distribution in (\ref{convdis}) is equal to $H(Y_2|X_2=\bar{x}_2, U)$, respectively $H(Y_2|X_2=\bar{x}_2)$, evaluated with the distribution in (\ref{achdis}). Furthermore, evaluated with the distribution in (\ref{achdis}),
\begin{align}
H(Y_2|X_2,U)&=\sum_{x_2,u}p^*(x_2)p(u)H(Y_2|X_2=x_2,U=u)\\
&=\sum_{x_2,u}p^*(x_2)p(u) H(Y_2|X_2=\bar{x}_2, U=u) \label{nequal1}\\
&=H(Y_2|X_2=\bar{x}_2,U),
\end{align}
We also have
\begin{align}
H(Y_2|X_2)=\sum_{x_2}p^*(x_2)H(Y_2|X_2=x_2)=H(Y_2|X_2=\bar{x}_2), \label{nequal1again}
\end{align}
where we obtain (\ref{nequal1}) and (\ref{nequal1again}) using Condition 1 with $n=1$.
Thus, we have proved that the region described by (\ref{specialupper1})-(\ref{specialupper3}) and that described by (\ref{achcon1})-(\ref{achcon3}) are the same when evaluated with the same $p(u)p(x_1|u)$.

Since the capacity region upper bound and the achievable rate region are taking the union of regions described by (\ref{specialupper1})-(\ref{specialupper3}) and (\ref{achcon1})-(\ref{achcon3}), respectively, over all $p(u)p(x_1|u)$, the achievable rate region and the capacity region upper bound coincide, yielding the capacity region.

Hence, the capacity region of the class of Z-interference channels that satisfies Conditions 1 and 2 is
\begin{align}
R_1 & \leq I(X_1;Y_1|U)+\min(I(U;Y_1), I(U;Y_2|X_2))\label{cap1}\\
R_2 & \leq \tau-H(Y_2|X_2,U) \label{cap2}\\
R_1+R_2 & \leq I(X_1;Y_1|U)+\tau-H(Y_2|X_2,U) \label{cap3}
\end{align}
for some distribution $p(u)p(x_1|u)$, where the mutual informations and entropies are evaluated with $p(u,x_1,x_2,y_1,y_2)=p(u)p(x_1|u)p^*(x_2)V_1(y_1|x_1)V_2(y_2|x_1,x_2)$. Using support lemma \cite[Lemma 3.4]{Csiszar:book}, without loss of generality, we may bound the cardinality of the auxiliary random variable $U$ as $|\mathcal{U}| \leq |\mathcal{X}_1|+1$.

\begin{figure}
\centering
\includegraphics[width=3.5in]{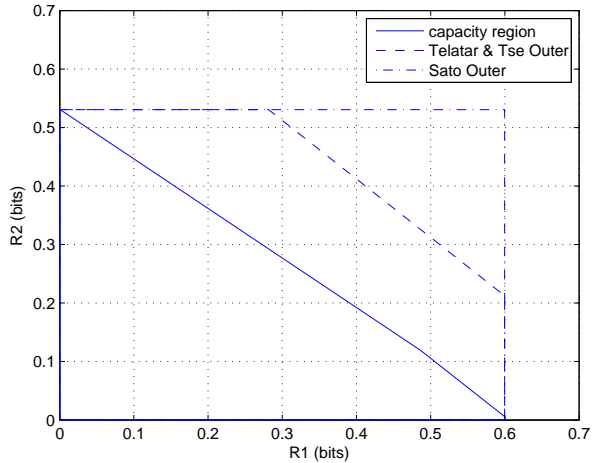}
\caption{Capacity region and known upper bounds for the Z-interference channel of Figure 1, where $p(y_1|x_1)$ describes a BEC with erasure probability .4 and $p(s|x_1)$ describes a BSC with crossover probability .1}\label{andreafig}
\end{figure}

In Figure \ref{andreafig}, we plot the capacity region and prior upper bounds by Telatar and Tse \cite{Telatar:2007} and by Sato \cite{Sato:1977} for the Z-interference channel based on the model illustrated in Figure \ref{eg}, where $\bar{p}(y_1|x_1)$ is a binary erasure channel (BEC) with erasure probability $0.4$ and $\bar{p}(s|x_1)$ is a binary symmetric channel (BSC) with crossover probability $0.1$. This Z-interference channel is based on the model illustrated in Figure \ref{eg}, and therefore satisfies both Conditions 1 and 2. Without loss of generality, by restricting the cardinality of the auxiliary random variable to be 3, we plot the capacity region as given by (\ref{cap1})-(\ref{cap3}). The dashed region is an inner bound of the upper bound given by \cite[Theorem 1]{Telatar:2007} where we have omitted time sharing. The dot-dash region is an upper bound given by \cite[Theorem 2]{Sato:1977} which outperforms the MAC, BC outer bounds in \cite{Sato:1977}. 
But for this channel model, the upper bound given by \cite[Theorem 2]{Sato:1977}  is simply equivalent to 
\begin{align}
\bigcup_{p(x_1),p(x_2)}\{(R_1,R_2)|R_1 \leq I(X_1;Y_1), \quad R_2 \leq I(X_2;Y_2|X_1)\},
\end{align}
which is the performance of both users where there is no interference. 
As can be seen, our new capacity region upper bound significantly tightens known upper bounds. More significantly, from the proofs in this section, we know that it equals the capacity region.

\section{Conclusions} \label{conclusions}
We have obtained a single-letter characterization for the capacity region of Z-interference channels that satisfy certain conditions. The two conditions are 1) that the performance of the second transmitter-receiver pair is invariant to the codeword locations of Transmitter 2 (i.e. it depends only on codeword distances), and 2) that regardless of the codebook choice of Transmitter 1, the input to Receiver 2 can always achieve maximum entropy.  The results are obtained through a new upper bound on the capacity region that exploits a technique by Korner and Marton. This new upper bound is shown to coincide with the Han/Kobayashi achievable rate region for channels satisfying the stated conditions. Thus, the Han/Kobayashi transmission strategy of superposition encoding and partial decoding is capacity-achieving for this class of channels. 

The upper bound also applies to channels satisfying the first but not the second condition. This bound significantly improves on existing upper bounds by Teletar/Tse and by Sato, as is illustrated with a numerical example. We also show that this upper bound coincides with the achievable rates at certain points on the capacity region, including the sum-rate point, for Z-interference channels satisfying the first condition. This upper bounding technique may also improve known bounds for other types of interference channels. 

\section{Appendix} \label{appendix}

\subsection{Proof of Theorem \ref{converse}}\label{converseproof}
Since the rate pair $(R_1, R_2)$ is achievable, there exist
two sequences of codebooks 1 and 2, denoted by $\mathcal{C}_1^n$ and $\mathcal{C}_2^n$, of rate $R_1$ and $R_2$, and probability of error less than $\epsilon_n$, where $\epsilon_n \rightarrow 0$ as $n \rightarrow \infty$.
Let $X_1^n$ and $X_2^n$ be uniformly distributed on codebooks 1 and 2, respectively. Let $Y_1^n$ be connected via $V_1^n$ to $X_1^n$, $Y_2^n$ be connected via $V_2^n$ to $X_1^n$ and $X_2^n$.

We start the converse with Fano's inequality \cite{Cover:book} and the data processing inequality \cite{Cover:book},
\begin{align}
nR_1=H(W_1) &=I(W_1;Y_1^n)+H(W_1|Y_1^n)\\
&\leq I(W_1;Y_1^n)+n \epsilon_n\\
&\leq I(X_1^n;Y_1^n)+n \epsilon_n\\
& = H(Y_1^n)-H(Y_1^n|X_1^n)+n \epsilon_n\\
&= H(Y_1^n)-\sum_{i=1}^n H(Y_{1i}|Y_1^{i-1}, X_1^n)+ n \epsilon_n\\
&=H(Y_1^n)-\sum_{i=1}^n H(Y_{1i}|X_{1i})+ n \epsilon_n \label{memoryless}
\end{align}
and (\ref{memoryless}) follows from the memoryless nature of $V_1^n$. We also have
\begin{align}
n R_2 =H(W_2) &= I(W_2;Y_2^n)+H(W_2|Y_2^n)\\
& \leq I(W_2;Y_2^n)+n\epsilon_n\\
& \leq I(X_2^n;Y_2^n)+n \epsilon_n\\
& = H(Y_2^n)-H(Y_2^n|X_2^n)+n \epsilon_n\\
& =\sum_{i=1}^n H(Y_{2i}|Y_2^{i-1})-H(Y_2^n|X_2^n)+n \epsilon_n\\
& \leq \sum_{i=1}^n H(Y_{2i})-H(Y_2^n|X_2^n)+n \epsilon_n \label{conditioning}
\end{align}
where (\ref{conditioning}) follows because conditioning reduces entropy.

Let us define another channel, $\hat{V}_2: \mathcal{X}_1 \rightarrow \mathcal{Y}_2$, as
\begin{align}
\hat{V}_2(t|x_1)=V_2(t|x_1, \bar{x}_2),
\end{align} 
where $\bar{x}_2$ is an arbitrary element in $\mathcal{X}_2$. Further, let us define another sequence of random variables, $T^n$, which is connected via $\hat{V}_2^n$, the memoryless channel $\hat{V}_2$ used $n$ times, to $X_1^n$, i.e., $T_i \rightarrow X_{1i} \rightarrow T_{\{i\}^c}, X_{1\{i\}^c}, X_2^n, Y_1^n, Y_2^n$. Also define $\bar{x}_2^n$ as the $n$-sequence with $\bar{x}_2$ repeated $n$ times. It is easy to see that
\begin{align}
H(Y_2^n|X_2^n)&=\sum_{x_2^n \in \mathcal{C}_2^n} \frac{1}{2^{nR_2}} H(Y_2^n|X_2^n=x_2^n)\\
&=H(Y_2^n|X_2^n=\bar{x}_2^n) \label{usecon1}\\
&= H(T^n), \label{useT}
\end{align}
where (\ref{usecon1}) follows from the fact that the channel under consideration satisfies Condition 1, and (\ref{useT}) follows from the definition of $T^n$. 

By applying Lemma \ref{nletterdifference}, we have
\begin{align}
H(T^n)-H(Y_1^n)=\sum_{i=1}^n \left( H(T_i|Y_1^{i-1}, T_{i+1}, T_{i+2}, \cdots, T_n)-H(Y_{1i}|Y_1^{i-1}, T_{i+1}, T_{i+2}, \cdots, T_n) \right).
\end{align}
Furthermore, since conditioning reduces entropy, we can write
\begin{align}
H(Y_1^n)& =\sum_{i=1}^n H(Y_{1i}|Y_1^{i-1}) \leq \sum_{i=1}^n H(Y_{1i})\\
H(T^n)&=\sum_{i=1}^n H(T_i|T^{i-1}) \leq \sum_{i=1}^n H(T_i)\\
H(Y_1^n)&=\sum_{i=1}^n H(Y_{1i}|Y_1^{i-1}) \geq \sum_{i=1}^n H(Y_{1i}|Y_{1}^{i-1}, T_{i+1}, T_{i+2}, \cdots, T_n).
\end{align}
Define the following auxiliary random variables,
\begin{align}
U_i=Y_{1}^{i-1}, T_{i+1}, T_{i+2}, \cdots, T_n, \qquad i=1,2,\cdots,n
\end{align}
Hence, we have
\begin{align}
\frac{1}{n} \left(H(T^n)-H(Y_1^n) \right)&=\frac{1}{n}\sum_{i=1}^n \left(H(T_i|U_i)-H(Y_{1i}|U_i) \right)\\
\frac{1}{n}\sum_{i=1}^n H(Y_{1i}|U_i) \leq \frac{1}{n}H(Y_1^n) & \leq \frac{1}{n} \sum_{i=1}^n H(Y_{1i})\\
\frac{1}{n} H(T^n)&\leq \frac{1}{n}\sum_{i=1}^n H(T_i).
\end{align}
Further define $Q$ as a random variable that is uniform on the set $\{1,2, \cdots, n\}$ and independent of everything else. Also, define the following auxiliary random variables:
\begin{align}
U=U_Q, \quad X_1=X_{1Q}, \quad X_2=X_{2Q}, \quad Y_1=Y_{1Q}, \quad T=T_Q, \quad Y_2=Y_{2Q}.
\end{align}
Then, we have
\begin{align}
\frac{1}{n} \left(H(T^n)-H(Y_1^n) \right)&=H(T|U, Q)-H(Y_1|U, Q) \label{korner1}\\
H(Y_1|U,Q) \leq \frac{1}{n}H(Y_1^n) &\leq H(Y_1|Q)  \label{korner2}\\
 \frac{1}{n}H(T^n) &\leq H(T|Q). \label{korner3}
\end{align}
Next, we prove some properties of the joint distribution of $Q$, $U$, $X_1$, $X_2$, $Y_1$, $T$ and $Y_2$. First, we have
\begin{align}
&\text{Pr}[X_2=x_2|X_1=x_1,U=u, Q=i]\\
&=\text{Pr}[X_{2i}=x_2|X_{1i}=x_1, U_i=u, Q=i]\\
&=\text{Pr}[X_{2i}=x_2|X_{1i}=x_1, (Y_{1}^{i-1}, T_{i+1}, T_{i+2}, \cdots, T_n)=u, Q=i]\\
&=\text{Pr}[X_{2i}=x_2|X_{1i}=x_1, (Y_{1}^{i-1}, T_{i+1}, T_{i+2}, \cdots, T_n)=u] \label{ridQfirst}\\
&=\text{Pr}[X_{2i}=x_2] \label{addx2}\\
&=\text{Pr}[X_2=x_2|Q=i], \label{ridQagain1}
\end{align}
where (\ref{ridQfirst}) and (\ref{ridQagain1}) are because $Q$ is independent of everything else, and therefore can be dropped from the conditioning, and (\ref{addx2}) follows because $X_{2i}$ is a function of message $W_2$, while $(X_{1i}, Y_{1}^{i-1}, T_{i+1}, T_{i+2}, \cdots, T_n )$ are functions of message $W_1$. Since $W_1$ and $W_2$ are independent, so is $X_{2i}$ and $(X_{1i}, Y_{1}^{i-1}, T_{i+1}, T_{i+2}, \cdots, T_n )$. We also have 
\begin{align}
&\text{Pr}[Y_1=y_1|X_1=x_1, X_2=x_2, U=u, Q=i]\\
&=\text{Pr}[Y_{1i}=y_1|X_{1i}=x_1, X_{2i}=x_2, Q=i, U_i=u]\\
&=\text{Pr}[Y_{1i}=y_1|X_{1i}=x_1, X_{2i}=x_2, Q=i, (Y_{1}^{i-1}, T_{i+1}, T_{i+2}, \cdots, T_n)=u]\\
&=\text{Pr}[Y_{1i}=y_1|X_{1i}=x_1, X_{2i}=x_2, (Y_{1}^{i-1}, T_{i+1}, T_{i+2}, \cdots, T_n)=u] \label{qind}\\
&=\text{Pr}[Y_{1i}=y_1|X_{1i}=x_1] \label{memorylesscheck1}\\
&=V_1(y_1|x_1), \label{dis1}
\end{align}
where (\ref{qind}) follows by the same reason as (\ref{ridQfirst}), and (\ref{memorylesscheck1}) follows from the memoryless nature of $V_1^n$.
We further have
\begin{align}
&\text{Pr}[T=t|X_1=x_1, X_2=x_2, U=u, Q=i, Y_1=y_1]\\
&=\text{Pr}[T_{i}=t|X_{1i}=x_1, X_{2i}=x_2, Q=i, U_i=u, Y_{1i}=y_1]\\
&=\text{Pr}[T_{i}=t|X_{1i}=x_1, X_{2i}=x_2, Q=i, (Y_{1}^{i-1}, T_{i+1}, T_{i+2}, \cdots, T_n)=u, Y_{1i}=y_1]\\
&=\text{Pr}[T_{i}=t|X_{1i}=x_1, X_{2i}=x_2, (Y_{1}^{i-1}, T_{i+1}, T_{i+2}, \cdots, T_n)=u, Y_{1i}=y_1] \label{qindagain}\\
&=\text{Pr}[T_{i}=t|X_{1i}=x_1] \label{memorylesscheck2}\\
&=V_2(t|x_1, \bar{x}_2), \label{dis2}
\end{align}
where (\ref{qindagain}) follows by the same reason as (\ref{ridQfirst}), and (\ref{memorylesscheck2}) follows from the definition of $T^n$, more specifically, the memoryless nature of $\hat{V}_2^n$ and the fact that $T_i \rightarrow X_{1i} \rightarrow T_{\{i\}^c}, X_{2i}, Y_1^n$, for $i=1,2, \cdots,n$.
Finally, we have
\begin{align}
&\text{Pr}[Y_2=y_2|X_1=x_1,X_2=x_2, U=u, Q=i, T=t, Y_1=y_1]\\
&= \text{Pr}[Y_{2i}=y_2|X_{1i}=x_1, X_{2i}=x_2, U_i=u, Q=i, T_i=t, Y_{1i}=y_1]\\
&= \text{Pr}[Y_{2i}=y_2|X_{1i}=x_1, X_{2i}=x_2, (Y_{1}^{i-1}, T_{i+1}, T_{i+2}, \cdots, T_n)=u, Q=i, T_i=t, Y_{1i}=y_1]\\
&=\text{Pr}[Y_{2i}=y_2|X_{1i}=x_1, X_{2i}=x_2, (Y_{1}^{i-1}, T_{i+1}, T_{i+2}, \cdots, T_n)=u, T_i=t, Y_{1i}=y_1] \label{ridQagain}\\
&=\text{Pr}[Y_{2i}=y_2|X_{1i}=x_1,X_{2i}=x_2] \label{memorylessy2}\\
&=V_2(y_2|x_1,x_2), \label{addy2final}
\end{align}
where (\ref{ridQagain}) follows by the same reason as (\ref{ridQfirst}), and (\ref{memorylessy2}) follows from the memoryless nature of $V_2$.
Thus, based on (\ref{ridQagain1}), (\ref{dis1}), (\ref{dis2}) and (\ref{addy2final}), we have shown that the joint distribution of $Q$, $U$, $X_1$, $X_2$, $Y_1$, $T$ and $Y_2$ satisfies
\begin{align}
p(q, u,x_1, x_2, y_1, t, y_2)&=p(q)p(u,x_1|q)p(x_2|q)V_1(y_1|x_1)V_2(t|x_1, \bar{x}_2)V_2(y_2|x_1,x_2). \label{distributionnan}
\end{align}
From (\ref{korner1})-(\ref{korner3}), we may conclude that there exists a number $\gamma$ that satisfies
\begin{align}
0 \leq \gamma \leq \min(I(Y_1;U|Q), I(T;U|Q)) \label{tconverse}
\end{align}
such that
\begin{align}
\frac{1}{n}H(T^n)=H(T|U, Q)+\gamma, \quad \frac{1}{n}H(Y_1^n)=H(Y_1|U, Q)+\gamma. \label{ttconverse}
\end{align}
From (\ref{dis1}), we see that
\begin{align}
\text{Pr}[Y_1=y_1|X_1=x_1, Q=i]=V_1(y_1|x_1),
\end{align}
which means 
\begin{align}
H(Y_1|X_1)=H(Y_1|X_1, Q)=\frac{1}{n}\sum_{i=1}^n H(Y_i|X_i). \label{glitch}
\end{align}
By combining (\ref{memoryless}), (\ref{conditioning}), (\ref{useT}), (\ref{distributionnan}), (\ref{tconverse}), (\ref{ttconverse}) and (\ref{glitch}), and allowing $n$ to approach infinity, we obtain the desired result.

\subsection{Proof of Theorem \ref{converse2}} \label{conversedifferent}
Because the distribution in Theorem \ref{converse} satisfies the Markov chain $(Q,U) \rightarrow X_1 \rightarrow Y_1$, we have
\begin{align}
H(Y_1|X_1)&=H(Y_1|X_1,U,Q). \label{swatch1}
\end{align}
Hence, comparing with the result of Theorem \ref{converse}, in order to prove this theorem, it is sufficient to show that for any $p(q)p(x_1,u|q)p(x_2|q)$, we have
\begin{align}
H(T|U,Q)&=H(Y_2|X_2,U,Q) \label{swatch2}\\
H(T|Q)&=H(Y_2|X_2,Q), \label{swatch3}
\end{align}
where entropies are evaluated with distribution 
\begin{align}
p(q, u,x_1, x_2, y_1,t, y_2)=p(q)p(x_1,u|q) p(x_2|q) \break V_1(y_1|x_1)V_2(t|x_1,\bar{x}_2)  V_2(y_2|x_1,x_2). \label{disappen}
\end{align} 
With the distribution in (\ref{disappen}), we have
\begin{align}
H(T|U=u,Q=q) &=H(Y_2|X_2=\bar{x}_2, U=u, Q=q) \label{converseadd1}\\
H(T|Q=q) & =H(Y_2|X_2=\bar{x}_2,Q=q).\label{converseadd2}
\end{align}
Moreover,
\begin{align}
H(T|U,Q)&=\sum_{q,u}p(q)p(u|q) H(T|U=u,Q=q) \label{converseadd3}\\
H(T|Q) & =\sum_{q}p(q) H(T|Q=q). \label{converseadd4}
\end{align}
In addition,
\begin{align}
H(Y_2|X_2,U,Q)&=\sum_{q,u,x_2} p(q)p(u|q)p(x_2|q) H(Y_2|X_2=x_2, U=u, Q=q)\\
&=\sum_{q, u,x_2} p(q)p(u|q)p(x_2|q) H(Y_2|X_2=\bar{x}_2, U=u, Q=q) \label{con1extra}\\
&=\sum_{q, u} p(q)p(u|q) H(Y_2|X_2=\bar{x}_2, U=u, Q=q) \sum_{x_2} p(x_2|q)\\
&=\sum_{q, u} p(q)p(u|q) H(Y_2|X_2=\bar{x}_2, U=u, Q=q) \label{converseadd5}
\end{align}
and
\begin{align}
H(Y_2|X_2,Q)&=\sum_{q,x_2} p(q)p(x_2|q) H(Y_2|X_2=x_2,Q=q)\\
&=\sum_{q,x_2} p(q)p(x_2|q) H(Y_2|X_2=\bar{x}_2,Q=q) \label{con1extra2}\\
&=\sum_{q} p(q) H(Y_2|X_2=\bar{x}_2,Q=q) \sum_{x_2} p(x_2|q)\\
&=\sum_{q} p(q) H(Y_2|X_2=\bar{x}_2,Q=q),\label{converseadd6}
\end{align}
where we obtain (\ref{con1extra}) and (\ref{con1extra2}) using Condition 1 with $n=1$. Using (\ref{converseadd1}), (\ref{converseadd2}), (\ref{converseadd3}), (\ref{converseadd4}), (\ref{converseadd5}) and (\ref{converseadd6}), we obtain (\ref{swatch2}) and (\ref{swatch3}), and thus have proved Theorem \ref{converse2}.

\bibliographystyle{unsrt}
\bibliography{ref}

\end{document}